# Harmonization among National Cyber Security and Cybercrime Response Organizations: New Challenges of Cybercrime

Yunsik Jake Jang[1]

Bo young Lim[2]

## Abstract

This presentation will discuss the need for national-level organizational strategies to effectively combat cyber security threats and cybercrime. In many countries, new agencies have been established and/or new roles have been allotted to existing agencies to cope with the needs for cyber security or fighting against cybercrime. The two pillars of organizational structure and functions (i.e., security vs. law enforcement) have given new challenges to us, especially in the context of traditional criminal justice system. To illustrate the challenges, a case study examining the responses to major security incidents followed by nationwide debates and remarkable organizational changes in Korea will be given.

## I. Introduction

The world today has never been more connected in human history. Owing to the remarkable progress people made in information technology, which was marked by the advent of the Internet, it became possible to overcome the barriers of time and space in the cyberspace. Now it takes only a few seconds for people to communicate with each other from halfway around the world, and regional events can have global implications. In this context, Asia's influence on world affairs in general is growing rapidly and will continue to grow in the next following decades. Asia boasts the largest population 'offline'; it continues its dominance 'online' as well, accounting for nearly 45% of all Internet users in the world[3]. Still this number is anticipated to grow further largely due to the increase of Internet users in China and India[4].

However, flip the coin; what you find is not a utopian 'small world', but a new space of opportunities for the 'bad guys,' i.e., criminals. What we see is a flood of news reports and publications dealing with phishing, hacking incidents, DoS attacks and viruses targeting individuals, corporations and government sites, which are all united under the word most frequently used by the media, 'cybercrime'. One of the most general assumptions on cybercrime is that as the number of Internet users increases, so will the number of cybercrime cases. This is usually drawn from the theory that crime follows opportunity, which has become established wisdom in criminology. Grabosky and Smith (1998) explains this by referring to a notorious American bank robber who, when asked why he persisted in robbing banks, replied: "Because that's where the money is." Linking this flow of logic to the significance Asia possesses in terms of Internet users, one might grimly conclude that Asia will have to face a dark future in cyberspace, having the biggest number of cybercrime cases than anywhere in the world.

---

[1] Professor, Department of Police Science, Korea National Police University

[2] M.A., Korea National Police Agency, Ph.D student at Korea University (Department of Law)

[3] Internet world stats, http://www.internetworldstats.com/stats.htm (Last accessed: July 28, 2012)

[4] Aguiar, M., Boutenko, V., Michael, D., Rastogi, V., Subramanian, A., & Zhou, Y. (2010). The Internet's New Billion: Digital Consumers in Brazil, Russia, India, China, and Indonesia. Retrieved from http://www.bcg.com/documents/file58645.pdf

But is it so? For criminologists, this is not a simple problem. First of all, to confirm whether the number of cybercrime really increases as the number of Internet users grew, reliable empirical data about criminal trends should be provided. However, in order to accumulate such data, an agreement on a general definition of cybercrime and standard methods to measure cybercrime should be reached. Secondly, unlike 'traditional' crimes, cybercrime has many overlapping areas that tend to blur the existing boundaries between the public police and other governmental/private organizations. This is a reason why many organizations and functions other than the public police are involved in dealing with cybercrime issues. Due to this nature of cybercrime, information about cybercrime does not flow through a single channel and is distributed among many relevant agencies[5]. Solving these problems is essential both in analyzing cybercrime from the criminological point of view and also improving the national-level organizational strategies to effectively combat cyber security threats and cybercrime.

This presentation aims to serve as a basis for solving this problem from a holistic view, focusing on the harmonization among relevant organizations and functions. In doing so, it will point out some practical problems and challenges that arise from the imbalance among different approaches to cybercrime by illustrating the case of South Korea. Finally, the presentation will conclude by giving some recommendations for national and international harmonization of organizational structure and its functions to improve global cooperation which is essential to protect the citizens from the borderless threats of cybercrime.

## II. The Impact of Cybercrime on Criminology and Criminal Justice
### i. Defining 'Cybercrime'

Defining the concept of 'cybercrime' is a critical matter; yet it is a difficult one which has huge impacts on not only legal issues such as the scope of jurisdiction but also on practical research designed to assess the impact of cybercrime on the society in general. Since policymaking relies heavily on the reliability of the information provided on the issue, the definition of cybercrime serves as a basis of quantitative measurement and qualitative classification. However, despite numerous attempts to define 'cybercrime' and classify it into categories, the following statement, which is part of the conclusion of a 1995 United Nations report that attempted to define 'computer crime,' still holds true for the concept of cybercrime as well:

> "There is no doubt among the authors and experts who have attempted to arrive at definitions of computer crime that the phenomenon exists. However, the definitions that have been produced tend to relate to the

---

[5] Wall, D. S. (2007). *Cybercrime: The Transformation of Crime in the Information Age*: Polity Press. p. 17. For more information on 'policing' the cyberspace and preventing cybercrime, see chapters 8 and 9 (pp. 157-206).

study for which they were written. [...] A global definition of computer crime has not been achieved; rather, functional definitions have been the norm.[6]"

The Council of Europe's Convention of Cybercrime, which is to date the only international convention related to the issue that is legally binding, uses 'cybercrime' as an umbrella term to refer to an array of criminal activity including the following: offenses against computer data and systems, computer-related offenses, content offenses, and copyright offenses[7].

Nevertheless, functional definitions are not flexible enough to embrace new types of cybercrime; given the significance and difficulty of defining cybercrime, it would be more reasonable to extend the discussion to consider the overall 'digitalization' of crime itself, rather than to confine the issue to looking for a separate way to cope with a new type of crime. According to Wall (2007), the term 'cybercrime' itself is fairly meaningless since it was largely an invention of the media and thus has been used metaphorically and emotionally rather than scientifically or legally. Rather, he argues that "the term [cybercrime] has a greater meaning if we construct it in terms of the transformation of criminal or harmful behavior by networked technology, rather than simply the behavior itself.[8]"

Although the term 'cybercrime' is indiscriminately used in the media, regardless of its type, it does not take long to realize that cybercrime is not a specific, single type of new crime. It is rather a set of various forms of crime that are related to the cyberspace one way or another that awaits comprehensive explanations and measures. Therefore, in devising a better strategy to effectively combat cybercrime, first it would be meaningful to explore how networked technology changed criminal behavior and the landscape of criminal justice in policing the cyberspace.

## ii. Criminological Explanations and the Need for Reliable Measurement

There have been many attempts made in the field of criminology to explain the transformation network technology brought to criminal behavior. In the United States, trends in violent and property crime shows that throughout the 1970s and 1980s the crime rate changed in proportion to the fluctuation of the youth population; as the volume of the youth population grew, so did the crime rate. However, starting from the mid 1990s, even though the youth population grew, the crime rate continued to fall[9].

---

[6] United Nations Manual on the Prevention and Control of Computer-Related Crime. (1995) *International Review of Criminal Policy*: United Nations.

[7] Jaishankar, K. (2008). Space Transition Theory of Cyber Crimes. In F. Schmalleger & M. Pittaro (Eds.), *Crimes of the internet*: Prentice Hall. p. 286

[8] Wall (2007). p. 10

[9] Schafer, J. A. Policing 2020: Exploring the Future of Crime, Communities, and Policing: Police Futurists International. p. 111. Retrieved from http://futuresworkinggroup.cos.ucf.edu/publications/Policing2020.pdf

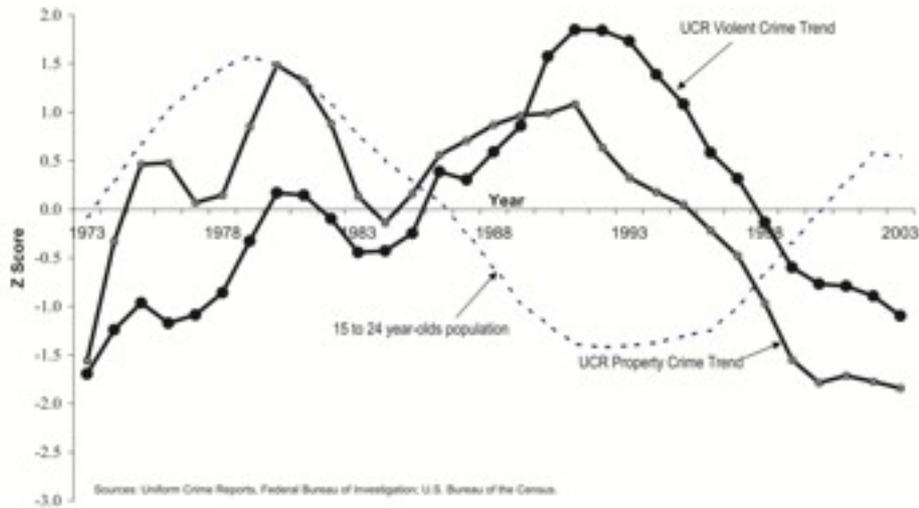

[Figure 1] U.S. Crime Trend, 1973-2003

  Ouimet (2008) identifies Internet usage as the main reason for this unexpected change in crime trends. Based on Cohen and Felson's Routine Activities Theory, he argues that since Netscape and Internet Explorer became standard in operating systems in 1994 and 1995, the Internet has brought a transformation in the daily routines of people, increased the possibility of crime detection because of the almost permanent traces, and provided people with information that has crime protection potential, and therefore contributed to the significant decrease of crime during the 1990s[10].

  While this may hold true for the sudden change in non-cybercrime trends of the United States in the 1990s, many questions still remain unanswered or inadequately addressed. For example, does this mean that the arrival of networked technology cause a decline in the overall crime rate? In contrast to the decrease of violent crime in the 1990s, complaints on all kinds of cybercrime have been showing a sharp increase, and the Internet Crime Complaint Center (IC3) now receives more complaints in a single month than it received in its first six months[11].

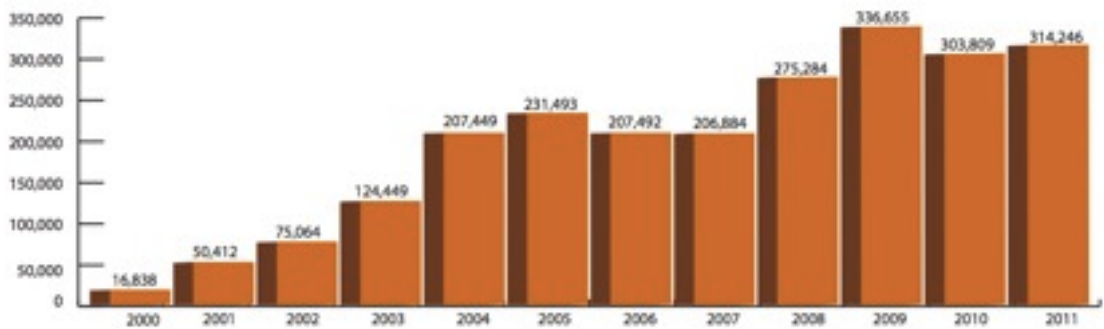

[Figure 2] Internet Crime Complaints, 2000-2011

---

[10] Ouimet, M. (2008). Internet and Crime Trends. In F. Schmalleger & M. Pittaro (Eds.), *Crimes of the Internet*: Prentice Hall.

[11] 2011 Internet Crime Report. (2011): Internet Crime Complaint Center. p. 7.

Then, in this case, can it be that the decline of crimes in the real world was actually a natural corollary of the crime transition from the physical space to cyberspace? Or can it be that the Internet rather brings an increase in crime rate by providing new opportunities to meet strangers in the cyberspace and plot a crime in the real world? Ouimet's explanation may account for the changes in crime trend of the United States during the 1990s, but it cannot be generalized to explain the changes in other Internet-using countries which show very different results regarding the impact of Internet usage on crime rates. This illustrates the need for a more comprehensive and generalizable theory for explaining cybercrime.

In fact, many attempts have been made to explain certain types of cybercrime with existing theories of criminology. But as cybercrime is a set of various types of crimes and does not clearly fit into the assumptions made by criminology, it is critical yet difficult to find a way to explain the phenomenon as a whole. In order to better understand the impacts of cybercrime Jaishankar (2008) has tried to provide an overall explanation for the phenomenon of cybercrime with his 'Space Transition Theory[12],' which postulates the following:

1. Persons, with repressed criminal behavior ( in the physical space) have a propensity to commit crime in cyberspace, which, otherwise they would not commit in physical space, due to their status and position.
2. Identity, Flexibility, Dissociative Anonymity and lack of deterrence factor in the cyberspace provides the offenders the choice to commit cybercrime.
3. Criminal behavior of offenders in cyberspace is likely to be imported to Physical space which, in physical space may be exported to cyberspace as well.
4. Intermittent ventures of offenders in to the cyberspace and the dynamic spatio-temporal nature of cyberspace provide the chance to escape.
5. (a) Strangers are likely to unite together in cyberspace to commit crime in the physical space. (b) Associates of physical space are likely to unite to commit crime in cyberspace.
6. Persons from closed society are more likely to commit crimes in cyberspace than persons from open society.
7. The conflict of Norms and Values of Physical Space with the Norms and Values of cyberspace may lead to cybercrimes.

Despite of the absence of a consistent and comprehensive definition of cybercrime, as criminology has started viewing the emergence of cyberspace as a new locus of criminal activity[13], Jaishankar's attempt to take an overall look at cybercrime is meaningful. However, although his theory seems to be persuasive, there is a need to test whether the above statements actually explains cyber-criminal activities. Perhaps the most general way to do this, as in verifying hypotheses in other fields as well, is by using quantitative data such as statistics.

---

[12] Jaishankar, K. (2008). Space Transition Theory of Cyber Crimes. In F. Schmalleger & M. Pittaro (Eds.), *Crimes of the internet*: Prentice Hall.

[13] Jaishankar, K. (2007). Establishing a Theory of Cyber Crimes. *International Journal of Cyber Criminology, 1*(2).

## iii. Changing Landscape of Criminal Justice in Policing the Cyberspace

The most generally used statistics in analyzing criminal behavior usually is that from public police departments, since the information of most traditional crimes would be reported to the police as that is where the case is to be resolved. This was possible because the criminal, victim, and the place where the case occurred were usually located in the same place or restricted to a certain area, and thus crime tended to be locally or nationally defined. Wall (2007) explains how cybercrimes made a change to this landscape. First, cybercrime contributed to "radical changes in the organization of crime and the division of criminal labor, and to changes in the scope of criminal opportunity[14]." This had a huge impact upon the existing localized criminal justice landscape by introducing a "new global dimension to the relationship between police, technology and the public[15]". Unlike traditional types of crime, networked technology makes it possible for cybercrime to overcome geographical constraints, making its reach global and inter-jurisdictional. According to Wall, in spite of various procedural and organizational responses, cybercrime poses a number of challenges to the traditional policing paradigm: *de minimism* ('the law does not deal with trifles'), which causes cybercrime to fall outside the traditional Peelian paradigm of policing dangerousness; *nullum crimen* (no crime without law) disparities in inter-jurisdictional cases; jurisdictional disparities; difficulties can arise when non-routine activities such as cross-border investigations happen, since the public police tends to be based upon local and 'routinized' practices; and under-reporting[16]. Consequently, cybercrime clearly falls outside the traditional purview of public police and causes under-reporting of cybercrime victimization to the police:

> "Put simply, relatively few Internet-related crimes are reported to the police because most are resolved elsewhere by the victims themselves, or by the panoply of other types of organizations or groups involved in the regulation of behavior in cyberspace.[17]"

Due to these reasons, Wall argues that the role of the public police in policing cybercrime should be understood as a small part of the networks of security within the cyberspace, and that the public police needs to forge new relationships with the other nodes within the network[18]. This argument is especially noteworthy, not only because of the distributed nature of cybercrime, but also because in reality there are many interest groups who play a role in policing cybercrimes other than the public police, either private organizations or governmental non-police organizations. As this is a relatively recent issue, public police in many countries are undergoing a transition period and are exploring how to forge partnerships with other organizations or agencies.

---

[14] Wall, D. S. (2007). *Cybercrime: The Transformation of Crime in the Information Age*: Polity Press. p. 39.

[15] Ibid., p. 160.

[16] See Ibid., pp. 161-166 for more information.

[17] Ibid., p. 166.

[18] Wall, D. S. (2011). Policing Cybercrimes: Situating the Public Police in Networks of Security Within Cyberspace. *Police Practice & Research: An International Journal, 8*(2), 183-205.

What this change in the landscape of criminal justice implies is that unlike the information of traditional crimes that usually flows through a single channel, the information of cybercrime is distributed among many private and public entities, so without adequate networking and collaboration among these organizations it is difficult to establish a comprehensive national, or even international, strategy to combat cybercrime. As Wall himself mentions, "We need reliable information about criminal trends not only to make sense of 'the problem', but also to act as a key driver of (criminal justice) policy reform and resource allocation within relevant agencies[19]. (...) The key issue here is about whether reliable information flows freely to form reliable viewpoints[20]." This account makes it clear that cooperation among various organizations with different functions is critical to grasp the situation of cybercrime and establish effective strategies to deal with it by enabling the provision of reliable information. From a long-term perspective, this is certain to have a bigger impact on the development of cyber criminology or criminology in general, both in verifying new theories that purport to explain cybercrime and in assisting the academic field to understand cybercrime more comprehensively.

But before arguing that multi-agency cross-sector partnerships are necessary for the development of criminology and more effective policies to combat cybercrime, taking a look at how the organizations involved in this issue are structured in reality would help to find more specific and feasible solution, since there exists a huge difference among states in the way the relevant entities are organized.

## III. Approaches on Cyber Attacks: Security vs. Law Enforcement
### i. The Two Approaches on Cyber Attacks

To date, many states have placed 'combating cybercrime' on the national agenda, and cybercrime has become an important issue at the global level as well. Yet not many countries have come up with an effective way to harmonize the functions of existing organizations and government agencies to better cope with the issue. Apart from the lack of a standard definition of cybercrime and the difficulty in measuring it, the absence of a holistic approach in viewing the issue of cybercrime stands as the main obstacle in establishing a comprehensive national strategy for cybercrime.

Currently there are two main approaches which is common at both the national and international level when viewing cyber attacks: a security-oriented approach, which first originated from viewing the threats from cyberspace as a technical threat to national security; and a law enforcement approach, which approaches the issue from a criminal justice point of view. The former has a tendency to focus on deterrence and prevention, while the latter emphasizes on investigation and attribution. This point is well identified in Maurer (2011)'s work. Maurer distinguished two streams of discourse in the United Nations on cyber security issues: the politico-military stream, which mainly has to do with cyber arms race; and the economic stream focusing on the criminal use of information technologies[21].

---

[19] Ibid., p. 13.

[20] Ibid., p. 23.

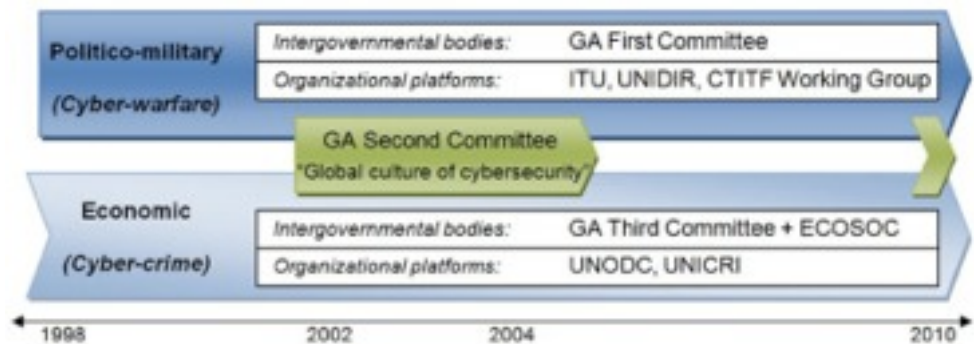

[Picture 1] Two Streams Model of Cyber Norm Emergence in the United Nations

While this Two Streams Model illustrates how international norms of cyber security is currently emerging in the United Nations and related organizations, it is nevertheless a result of close observation of the activities related to cyber norms emergence and thus has no normative implications for how the issue should be dealt with in reality. Just as cybercrime changed the landscape of criminal justice as mentioned above, the bigger picture of cyber attacks in general has also blurred the traditional boundaries that existed between crime and warfare.

Traditionally, regulation and deterrence of crime belongs to the purview of criminal justice and the perpetrator is punished by the law enforcement, which is mostly based on domestic law. On the other hand, warfare is conflict between states and follows the rules of international law. When physical attack was the only possible form of aggression, this dual approach and response was quite reasonable. But since stories of cyber attacks started to make the front page in all newspapers, it is becoming all the more difficult to conceptually distinguish between warfare and crime and decide how to resolve a certain case.

## ii. The Problem of Demarcation and Attribution of Cyber Attacks

Nye (2010) has classified four major cyber threats to national security as economic espionage, crime, cyber war, and cyber terrorism[22]. These concepts are widely used, despite the fact that there is no universal definition for these concepts, making demarcation among them difficult. Nevertheless, basically all four concepts have one thing in common: they are all subject to the regulation of law. Cyber war should be ruled by international law, while cyber terrorism, cybercrime and economic espionage can be categorized as crime and is subject to criminal law. Since there is no fully developed international practice or norm regarding cyber war to date and both economic espionage and cyber terrorism can be classified as a type of cybercrime, there is no point in distinguishing between the type of legislation applied to each of the four attacks. Rather, demarcation has a bigger meaning in practical terms such as deciding the division of labor among relevant organizations and agencies in response to such incidents.

---

[21] Maurer, T. (2011). Cyber Norm Emergence at the United Nations - An Analysis of the UN's Activities Regarding Cyber-Security? : Belfer Center for Science and International Affairs, Harvard Kennedy School.

[22] Nye, J. S. (2010). Cyber Power: Harvard Kennedy School Belfer Center for Science and International Affairs.

Yet, even if one day we reach a point where the demarcation problem of cyber attacks based on standard definitions is resolved, due to the nature of the cyberspace it is difficult to attribute a certain attack to one of the four types of cyber attack. Technically this is because it is difficult to trace the attacker, and easier for the attacker to disguise or detour, Other than technical reasons there are legal problems in tracing information located overseas and difficulties in acquiring voluntary cooperation from foreign counterparts. Since a precise method for attribution has yet to be developed, the attribution process should be based upon several elements, such as the type of attack, subject, period of time, the size and impact of the attack, target of attack, and motive, put together. However, it is very difficult to collect all these information perfectly due to the distributed nature of cyber attacks and the lack of partnership among relevant organizations at the current stage.

Regarding criminal investigation, the discussion on the role in attribution is especially important. This point was already made by the United States in 2003 in The National Strategy to Secure Cyberspace:

> "Law enforcement and the national security community play a critical role in preventing attacks in cyberspace. **Law enforcement plays the central role in attributing an attack through the exercise of criminal justice authorities**. Many cyber-based attacks are crimes. As a result the Justice Department's Computer Crime and Intellectual Property Section, the FBI's Cyber Division, and the U.S. Secret Service all play a central role in apprehending and swiftly bringing to justice the responsible individuals. (...) Ideally, an investigation, arrest, and prosecution of the perpetrators, or a diplomatic response in the case of a state-sponsored action, will follow such an incident[23]."

It is a natural corollary for the law enforcement to play the central role in attribution since legal authority prescribed by the criminal justice procedure is necessary in identifying the origin and thus attributing a cyber attack case to a certain type of attack, and because it is generally allowed to take a criminal justice approach regardless of the type of attack. Under normal situations, a case should be first attributed to a certain type through investigation led by law enforcement agencies in order to decide how to respond to it depending on its type of attack.

However, the overall structure of cyber attacks response differs from one state to another, and some countries have cyber crisis management systems that put defense and damage restoration above cyber attack attribution. By doing so, it becomes extremely difficult not only in the attribution of a certain cyber attack, but also in generating information about the "changing nature of threats, the characteristics and methodologies of threats, and emerging threat idiosyncrasies for the purpose of developing response strategies and reallocating resources, as necessary, to accomplish effective prevention[24]," which is critical for both the academic circle of criminology and decision-makers who establish national strategies for the cyberspace. A good example is the case of South Korea.

---

[23] The National Strategy to Secure Cyberspace. (2003): The White House. p. 43.

[24] Carter, D. L., & Schafer, J. A. The Future of Law Enforcement Intelligence. In J. A. Schafer (Ed.), *Policing 2020: Exploring the Future of Crime, Communities, and Policing*: Police Futurists International. p. 235.

# III. Case Study: South Korea's National Cybersecurity Structure

## i. Current National Cyber Security Structure of South Korea

The 'information security policy' was first implemented in the mid 1990s as a part of 'informatization' strategy. In 1996, the Korean Information Security Center (currently Korea Internet & Security Agency) was established, based on the Act on Expansion and Dissemination and Promotion of Utilization of Information System (currently the Act on Promotion of Information and Communications Network Utilization and Information Protection, etc., which works as a mother law for regulating most of the misconducts on cyberspace). In 2000, triggered by the Mafia boy's DDoS attacks and global media attention, the Cyber Terror Response Center (CTRC) of Korea National Police Agency was established. Since then, major security incidents have provoked changes in the national cyber security strategy and organizational structures. For example, in the aftermath of the 2003. 1. 25. Slammer Worm incident, also known as 1·25 Internet Disaster in South Korea, several new organizations were launched following national concerns. The National Cyber Security Center in the National Intelligence Service (NIS) became responsible for national cyber security in the public sector, while the Ministry of National Defense (MND) set up the Information Warfare Response Center to handle cyber warfare, and the Korea Internet Security Center (KISA) launched the Korean Internet Incident Response Center which became responsible for national cyber security in the private sector. This set of alphabet soup triggered the initial national cyber security framework. After the 2004 National Espionage case, in which the investigation started from a civillian's report about a spoofed email and some sort of attribution was successful, the government announced the Presidential Directive for National Cyber Security Regulation and the Presidential Directive for National Crisis Management, both of which are controlled by the National Security Council (NSC) secretariat. Following the 2009 7·7 DDoS attack, which was traced back to the origin of North Korea by the CTRC, in August the Comprehensive Countermeasure for Cyber Crisis was released. This initiative in effect eliminated the roles of the NSC and reinforced the responsibilities of the NIS. In 2010, the Cyber Command was established in the MND.

Centering on security incident response, currently there are three national frameworks: Critical Information Infrastructure (CII) protection framework, which is based on the Act on the Protection of Information and Communications Infrastructure of 2001 and is chaired by the Minister of Prime Minister Office (PMO); National Cybersecurity Regulation (NCR), which is based on the Presidential Directive for National Cyber Security Regulation and is chaired by the head of NIS, is applied to all governmental information systems except for CII; and Security Incident Response in Private Sectors, based on the Act on Promotion of Information and Communications Network Utilization and Information Protection.

## ii. Problems of the Current Structure

### (i) National Cybersecurity Regulation (NCR)

Among these three frameworks, the NCR represents the overwhelming control of the National Intelligence Service over the governmental information system.

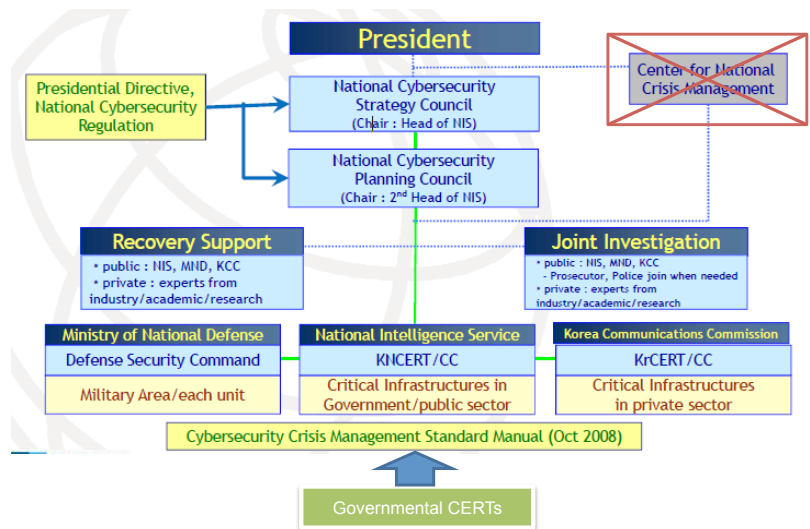

[Picture 2] National Cybersecurity Regulation

Normally governmental CERTs are in charge of the monitoring, and relevant information is transferred to the NIS through networks. When an incident is detected, it is required to report to the NIS. The inspections are carried out by governmental agencies if it seems to be a 'trivial' case, but by the NIS if is a 'serious' one. When national security is concerned, it is mandatory for law enforcement agencies to inform the NIS of the criminal case, whereas it is optional for the NIS to notify law enforcement agencies of the case if there is any criminal suspicion since. As most cyber incidents are crimes under the existing law, this optional clause could be in violation of the Criminal Procedure Act, which states that "when a public official in the course of his/her duty believes that an offense has been committed, he shall lodge an accusation" (Art. 234 (2)). Moreover, this regulation also seriously undermines the basis for cybercrime measurement. Since 2009, the NIS stopped announcing the statistics of incidents against governmental information systems. As most cases handled by the NIS does not undergo police investigation, the nondisclosure of statistics severely flaws the current cybercrime statistics announced by the police.

In addition, recently the government has announced the National Cyber Security Master Plan in August 2011, which designated the Director of the NIS to oversee and coordinate national cybersecurity-related policy and management. This can be very dangerous, since national intelligence services act relatively free from legal restrictions due to the nature of its mission. When NIS operations are confined to providing national security intelligence to "support decision" making, the low legal threshold applied to the NIS usually does not matter. However, giving the NIS the authority to "coordinate" policy without any legal restrictions threatens the balance of power, which is one of the principles of democracy.

(ii) Security Incident Response in Private Sectors

The framework for Security Incident Response in Private Sectors also shows a simliar problem of over-centralization and lack of harmonization among relevant organizations. Article 48-3 of The Act on Promotion of Information and Communications Network Utilization and Information Protection, on which this framework is based, stipulates obligation to immediately report incidents against 1) a provider of information and communications services or; 2) A business operator of clustered information and communications facilities. Furthermore, in Article 48-4 (5):

> "The Korea Communications Commission or the joint private-public investigation team **shall not use the information** learned through the data submitted and the investigation conducted in accordance with paragraph (4) **for any purposes other than analysis of causes of the intrusion and preparation of countermeasures, and shall destroy it immediately after the analysis of causes is completed**."

Although the role of law enforcement is indispensable due to the legal authority required in the criminal procedure in attribution since it is necessary to identify the origin of the attack, nowhere does the Act clearly mention the need to collaborate with law enforcement agencies. Moreover, Article 48-4 (5) further impedes attempts to measure cybercrime. To borrow the words of Wall, this not only preclude relevant agencies from acquiring reliable information about criminal trends, but also severely undermines the resources policymakers can refer to when making decisions for a better national cyber security policy and resource allocation within relevant agencies.

(iii) Failure to Recognize the Role of Law Enforcement and Cybercrime Measurement

While the National Cybersecurity Regulation grants excessive access and control to information of cyber incidence to the intelligence service, the Act on Promotion of Information and Communications Network Utilization and Information Protection restricts the use of information other than the analysis of cause of the intrusion and preparation of countermeasures, and requires the immediate destruction of information after such analysis is completed, thereby making the measurement of cybercrime difficult and also impedes criminal investigation on cyber incidents.

Both frameworks have little in mind of the nature of cyber attacks, i.e., the fact that most types of cyber attacks can and should be dealt according to the criminal law, and lacks an overall view of the problem, thereby granting excessive authority to either the intelligence service or the security community while failing to recognize the need to work with the law enforcement. This is a stark contrast compared to the incident response system of the United States, which is frequently referred to as a model cyber security framework. According to the Federal Information Security Management Act (FISMA), federal agencies must report incidents to the United States Computer Emergency Readiness Team (US-CERT) and the details are guided by the Computer Security Incident Handling Guide, which mentions the role of law enforcement as follows:

> "One reason that many security-related incidents do not result in convictions is that organizations do not properly contact law enforcement. (...) The incident response team should become acquainted with its various law enforcement representatives before an incident occurs to discuss conditions under which incidents should be reported to them, how the reporting should be performed, what evidence should be collected, and how it should be collected. Law enforcement should be contacted through designated individuals in a manner consistent with the requirements of the law and the organization's procedures. Many organizations prefer to appoint one incident response team member as the primary POC with law enforcement. This person should be familiar with the reporting procedures for all relevant law enforcement agencies and well prepared to recommend which agency, if any, should be contacted[25]."

Not to mention that this failure to recognize the importance of law enforcement in incident management, it also seriously undermines the grounds for reliable measurement of cybercrime, either by giving the intelligence excessive authority in handling relevant information while paying insufficient attention to the role law enforcement plays in attribution, or by restricting the use of information of private sector security incidents to the analysis of causes of the intrusion and preparation of countermeasures, thereby blocking the information flow to law enforcement agencies for criminal investigation.

## iv. Need for Harmonization among Organizations

Taking a look at the big picture of national security, the problems mentioned above seem to originate from the lack of consistency in the national cyber security strategy and organizational structures. As mentioned earlier, the current national cyber security structure was subject to changes every time a major security incident occurred. The words of Schnier (2000) are especially meaningful in pointing this out:

> "The policy [an overall strategy] is what ties everything together. (...) [a security policy should be built] based on the threat analysis, (...) The policy should outline who is responsible for what (implementation, enforcement, audit, review), what the basic network security policies are, and why they are the way they are. The last one is important; arbitrary policies brought down from on high with no explanation are likely to be ignored. A clear, concise, coherent, and consistent policy is more likely to be followed. (...) **In any case, the security policy needs to outline "why" and not "how"**[26]."

But the lack of consistency is not the root of the problem. Why new agencies were established every time an incident occurred is because South Korea does not have an overall national strategy based on inadequate threat analysis. This is why the current national cyber security structure of South Korea only explains "how" to deal with cyber incidents and which agency carries out what responsibilities, but does not tell "why" it has granted the NIS de facto control over incidents of governmental information networks, or "why" the information of private sector security incidents should be immediately destroyed after the analysis of cause is completed. Although the issue of cyber incident response may appear to be a complex issue, if anyone paid sufficient attention to the nature of cyber attacks and recognized the difficulty of attribution and measurement, law enforcement may have been playing an important role in detecting, attributing, and identifying the perpetrator.

---

[25] Scanfone, K., Grance, T., & Masone, K. (2008). Computer Security Incident Handling Guide *NIST Special Publication*: National Institute of Standards and Technology. p. 2-6.

[26] Schneier, B. (2000). *Secrets & Lies: Digital Security in a Networked World*. Indianapolis, Indiana: Wiley Publishing, Inc. pp. 308-309.

Nevertheless, "security is a process, not a product[27]." Although the current national cybersecurity structure of South Korea has many problems, they can be resolved by promoting harmonization among relevant organizations.

---

[27] Ibid., p. 84.